\documentclass[12pt,letterpaper]{article}
\usepackage{amsmath}
\usepackage{amssymb}
\usepackage{cite}
\usepackage{dcolumn} 
\usepackage{setspace}
\usepackage{array}
\def\brppp{{\mathbf{r}^{\prime\prime\prime}}}
\def\brpp{{\mathbf{r}^{\prime\prime}}}
\def\brp{{\mathbf{r}^{\prime}}}
\def\tp{{{t}^{\prime}}}

\def\tbr{{\tilde{\mathbf{r}}}}
\def\br{{\mathbf{r}}}

\def\bR{{\mathbf{R}}}
\def\bM{{\mathbf{M}}}

\def\bK{{\mathbf{K}}}

\def\d{{\mathrm{d}}}
\def\rhor{{\rho({\bf r})}}
\def\rhorp{{\rho({\bf r}^{\prime})}}
\def\rhoi{{\rho_I}}
\def\rhoj{{\rho_J}}
\def\rhoir{{\rho_I({\bf r})}}

\def\rhojrp{{\rho_J({\bf r}^{\prime})}}
\def\sumi{{\sum_I^{N_S}}}
\def\sumj{{\sum_J^{N_S}}}

\newcommand{\eqn}[1]{\mbox{Eq.\hspace{1pt}(\ref{#1})}}
\newcommand{\eqs}[2]{\mbox{Eqs.\hspace{1pt}(\ref{#1}--\ref{#2})}}
\newcommand{\pot}[1]{v_{\rm #1}}

\begin{document}

\begin{center}
\vspace*{1cm}
{\LARGE\bf On the Subsystem Formulation of Linear Response Time-Dependent DFT}\\[3ex]
{\large Michele Pavanello\footnote{E-mail: 
m.pavanello@rutgers.edu}}\\[2ex]
Department of Chemistry, Rutgers University,\\
Newark, NJ 07102, USA\\
\end{center}

\vfill

\begin{tabbing}
Date:   \quad\= \today \\
Status: \> Accepted by J. Chem. Phys. on May 3, 2013 \\
\end{tabbing}
\newpage
\begin{abstract}
A new and thorough derivation of linear-response subsystem TD-DFT is presented and analyzed in detail. Two equivalent derivations are presented and naturally yield self consistent subsystem TD-DFT equations. One derivation reduces to the subsystem TD-DFT formalism of Neugebauer [J. Neugebauer, {\it J. Chem. Phys.} {\bf 126}, 134116 (2007)]. The other yields Dyson type equations involving three types of subsystem response functions: coupled, uncoupled and Kohn--Sham. The Dyson type equations for subsystem TD-DFT are derived here for the first time.

The response function formalism reveals previously hidden qualities and complications of the subsystem formulation of TD-DFT compared with the regular TD-DFT of the supersystem. For example, analysis of the pole structure of the subsystem response functions shows that each function contains information about the electronic spectrum of the entire supersystem. In addition, comparison of the subsystem and supersystem response functions shows that, while the correlated response is subsystem additive, the Kohn--Sham response is not. Comparison with the non-subjective Partition DFT theory shows that this non-additivity is largely an artifact introduced by the subjective nature of the density partitioning in subsystem DFT.
\end{abstract}
\newpage
\section{Introduction}
When modeling systems that contain a large number of electrons, 
even the Kohn--Sham Density Functional Theory (KS-DFT) approach \cite{kohn1965} has its limits. In the past decade many approximations \cite{scus1999,goed1999,bowl2012} made it possible to massively reduce KS-DFT complexity for spatially extended molecules. However, the large pre-factor of such scaling laws left the calculation of most realistic, fully-solvated systems still prohibitive \cite{carl2002,burk2012}. 

Reducing the computational complexity of KS-DFT by partitioning the total electron density of a system into subsystem contributions has been an appealing idea since the early works of Gordon and Kim \cite{gord1972,kim1974}. However, the success of KS-DFT seemed to have rendered partitioning methods unnecessary. This is evident from the Quantum Chemistry literature of the 70s and 80s, where partitioning methods were frequent only to high-end wave function methods, and interactions between subsystems were treated with various types of perturbation theory \cite{jezi1982}. Despite two successful applications of density partitioning techniques, first by Senatore and Subbaswamy \cite{sen1986}, and then by Cortona \cite{cort1991}, revival of these methods is due to a paper by Wesolowski and Warshel published in 1993 \cite{weso1993}.
Presently, subsystem DFT is being developed by many research groups worldwide \cite{silv2012,hu2012,pava2012d,gome2012,hofe2012,lari2012,good2011a,neug2010a,iann2006}. Successful applications of subsystem DFT are reported for applications related to the ground state, such as analysis of electron densities \cite{fux2008}, and spin densities \cite{solo2012}; and for calculations of charge and excitation energy transfer parameters \cite{pava2012d,pava2011b,neug2010}; as well as for electronic spectra and molecular properties \cite{tecm2012,koni2012b,koni2011,neug2010a,bulo2008,jaco2006b}.

The time-dependent extension of susbsystem DFT has been pioneered by Casida and Wesolowski \cite{casi2004}. However, Neugebauer \cite{neug2007} is credited for deriving working equations for the solution of the subsystem time-dependent DFT (subsystem TD-DFT, hereafter) and for applying subsystem TD-DFT to determine excitation energies \cite{neug2007,neug2010a,koni2011}, charge/exciton couplings \cite{pava2012d,pava2011b,neug2010}, and molecular properties \cite{neug2009}. Similarly to subsystem DFT, subsystem TD-DFT is developed to take full advantage of the subsystem nature of the majority of real life systems. Solvated systems are a typical example of this. An early success story of subsystem TD-DFT is the calculation of solvatochromic shifts \cite{neug2005b}. More recently, the electronic spectra of light harvesting complexes model systems containing more that 1000 atoms has been calculated with this method \cite{koni2011}.

Linear-response TD-DFT has been formulated by many authors in many publications. This has frequently offered a chance to discuss its limitations and to offer possible solutions. This has not been the case for subsystem TD-DFT. Partly because this field is relatively new. 

The large body of work on subsystem DFT and TD-DFT shows their usefulness and importance. However, a work that aims at clarifying the relationship between subsystem and supersystem TD-DFT, and analyzing what the density partitioning does to the collective time-dependent response of the system is long overdue. This work, aims at filling this gap providing two derivations of subsystem TD-DFT. The new derivations are amenable to a deeper analysis and understanding of the theory of subsystem TD-DFT. For example, Dyson type equations for susbsystem TD-DFT are derived here for the first time.

This work is organized as follows. In the next section, a theoretical background is given on KS-DFT, subsystem DFT, and linear-response TD-DFT. In Section \ref{sect_sTD-DFT}, a rigorous derivation of linear-response subsystem TD-DFT is carried out. In Section \ref{sect_sTD-DFT2} an alternative derivation of subsystem TD-DFT is presented in terms of subsystem response functions. Section \ref{sect_supra} is devoted to the comparison of subsystem TD-DFT with TD-DFT of the supersystem. In Section \ref{sect_conc}, conclusions are drawn.
\section{Theoretical Background}
\subsection{Ground state DFT and subsystem DFT}
KS-DFT can be summarized by the following equation, known as the KS equation, in canonical form,
\begin{equation}
\label{ks1}
\left[ -\frac{1}{2} \nabla^2 + \pot{eff}(\br) \right] \phi_k(\br) = \varepsilon_k  \phi_k(\br),
\end{equation}
where $\pot{eff}$ is the effective potential that the one-particle KS orbitals, $\phi_k$, experience, and $\varepsilon_k$ are the KS orbital energies. The spin labels have been omitted for sake of clarity, as throughout this work only the spin restricted case is considered without loss of generality of the derivations. The electron density is simply $\rhor=2\sum_i^{\rm occ} \left| \phi_i(\br)\right|^2$.

The effective potential, $\pot{eff}$, is given by
\begin{equation}
\label{ks2}
\pot{eff}(\br)=\pot{appl}(\br)+\pot{eN}(\br)+\pot{Coul}(\br)+\pot{xc}(\br),
\end{equation}
with $\pot{appl}$ being an externally applied potential, $\pot{eN}$ the electron--nucleus attraction potential, $\pot{Coul}$ the Hartree potential, and $\pot{xc}$ the exchange--correlation (XC) potential\cite{kohn1965}. 

Subsystem DFT is based on the idea that an electronic (molecular) system can be more easily approached if it is partitioned into many smaller subsystems. In mathematical terms, this is done by partitioning the electron density as follows \cite{sen1986,cort1991}
\begin{equation}
\label{fde1}
\rhor=\sumi \rhoir,
\end{equation}
with $N_S$ being the total number of subsystems. 

The ultimate goal is to represent the subsystems as a set of $N$ coupled Kohn--Sham systems. Hence, the subsystem densities must be non-negative, must integrate to a preset number of electrons, i.e.\ $\int\rhoir\d\br=N_I$, and must be $v$-representable. In this context, it is perfectly legitimate to wonder what then constitutes a subsystem. The three requirements (constraints) mentioned above constitute the only theoretical prescription.  It is remarkable that this prescription does not invoke any real space partitioning. 

Therefore, subsystem densities can, in principle, strongly overlap and can be highly delocalized. In practical calculations, however, the subsystem densities are constructed from subsystem molecular orbitals which are expanded in terms of localized atomic orbitals, often centered to atoms belonging to only one molecular fragment in a system (monomer basis set). In practical calculations, the latter approximation and the use of local and semilocal non-additive kinetic energy functionals define the subsystems as non-covalently bound molecules.

Self consistent solution of the following coupled KS-like equations (also called KS equations with constrained electron density \cite{weso2006}) yield the set of subsystem KS orbitals, i.e.
\begin{equation}
\label{ks1}
\left[ -\frac{1}{2} \nabla^2 + \pot{eff}^I(\br) \right] \phi^I_k(\br) = \varepsilon^I_k  \phi^I_k(\br),\mathrm{~with~}I=1,\ldots,N_S
\end{equation}
with the effective subsystem potential given by
\begin{equation}
\label{sks}
\pot{eff}^I (\br)=\underbrace{\pot{appl} (\br)+\pot{eN}^I (\br) + \pot{Coul}^I(\br) +\pot{xc}^I (\br)}_{\mathrm{same~as~regular~KS-DFT}} + \pot{emb}^I (\br).
\end{equation}
In the above it is clear that if an applied potential, $\pot{appl}$, acts on the total system, every subsystem will experience that same potential.
In the so-called Frozen Density Embedding (FDE) formulation of subsystem DFT \cite{weso1993,weso2006}, the unknown potential above, $\pot{emb}$, is called embedding potential and is given by
\begin{align}
\label{emb}
\nonumber
\pot{emb}^I (\br)&=\sum_{J\neq I}^{N_S} \left[ \int \frac{\rhojrp}{|\br-\brp|} \d\brp - \sum_{\alpha\in J} \frac{Z_\alpha}{|\br-\bR_{\alpha}|} \right] + \\
&+\frac{\delta T_s[\rho]}{\delta\rhor}-\frac{\delta T_s[\rhoi]}{\delta\rhoir}+\frac{\delta E_{\rm xc}[\rho]}{\delta\rhor}-\frac{\delta E_{\rm xc}[\rhoi]}{\delta\rhoir}.
\end{align}

Throughout this work, ``subsystem DFT'' is used as a synonym of FDE. The density of the supersystem is thus found using \eqn{fde1} and \eqn{ks1} as $\rhor=2\sumi\sum_i^{{\rm occ}_I} \left| \phi^I_i(\br)\right|^2$.

\subsection{Linear-response TD-DFT}
The time-dependent KS equation,
\begin{equation}
\label{casi1}
\left[ -\frac{1}{2}\nabla^2 + \pot{eff}(\br,t) \right] \phi_k(\br,t)= i \frac{\partial \phi_k(\br,t)}{\partial t},
\end{equation}
relates the time dependent KS orbitals, $\phi_k(\br,t)$, and the correlated density, $\rho(\br,t)=2\sum_{k}^{\rm occ} |\phi_{k}(\br,t)|^2 $, with the externally applied, time-dependent perturbation, $\pot{appl}(\br,t)$. When starting from the ground state density, the time-dependent KS potential is defined according to \eqn{ks2} as
\begin{equation}
\label{casi2}
\pot{eff} (\br,t) = \pot{appl}(\br,t) + \pot{eN} (\br) + \pot{Coul} (\br,t) + \pot{xc} (\br,t),
\end{equation}
The applied potential constitutes the only time-dependent perturbation causing the density $\rho$ to become a time-dependent function \cite{casi2004,casi2012}. With the exception of $\pot{eN}$, which is considered static (the nuclei are assumed to be still in the time the perturbation is applied), the other potential terms part of the effective KS potential are dependent on time, but only as a result of the perturbation $\pot{appl}(\br,t)$ through their density dependence.

Linear-response TD-DFT is based on the assumption that the density response to the external weak perturbation is given by the following linear-response integral equations \cite{casi1995}
\begin{align}
\label{casi3a}
\delta \rho(\br,t)&=\int \chi(\br,\brp,t-\tp) \delta \pot{appl} (\brp,\tp)\d\brp \d\tp\\
\label{casi3b}
                                  &=\int \chi^0(\br,\brp,t-\tp) \delta \pot{eff} (\brp,\tp)\d\brp \d\tp,
\end{align}
where 
\begin{equation}
\label{casi4}
\delta \pot{eff}(\brp,t)=\delta \pot{appl}(\brp,t) + \delta \pot{ind} (\brp,t),
\end{equation}
The induced potential, $\delta \pot{ind}$, is expressed in linear-response as well, namely
\begin{equation}
\label{casi5}
\delta \pot{ind} (\brp,t)=\int_{\tp=t_0}^{\tp=t}\int{\left[ \frac{\delta(t-\tp)}{|\brp-\brpp|} + \frac{\delta  \pot{xc} (\brp,t)}{\delta\rho(\brpp,\tp)} \right] \delta\rho(\brpp,\tp)} \d\brpp \d\tp.
\end{equation}
The quantity $ \frac{\delta  \pot{xc} (\brp,t)}{\delta\rho(\brpp,\tp)}$ is called the XC kernel, $f_{\rm xc}(\brp,\brpp,t-\tp)$. The functions $\chi(\br,\brp,t-\tp)$ and $\chi^0(\br,\brp,t-\tp)$ are the correlated and the simplified KS response functions (or simply ``correlated response'' and ``KS response'', respectively). \eqn{casi3a} constitutes the definition of linear-response TD-DFT, and \eqn{casi3b} derives from \eqn{casi3a} from the Runge--Gross theorem (Theorem 4 of Ref.\ \cite{rung1984}).  

As it is more convenient to write the working equations in the frequency domain, by virtue of the convolution theorem, the above equation can be rewritten as
\begin{equation}\label{casi5w}
\delta \pot{ind} (\brp,\omega)=\int{\left[ \frac{1}{|\brp-\brpp|} + f_{\rm xc}(\br,\brp,\omega) \right] \delta\rho(\brpp,\omega)} \d\brpp
\end{equation}

For practical calculations, \eqn{casi3b} is the most important. This is because, in the adiabatic approximation, it involves quantities that can be extracted from the ground state KS system, such as the KS response function (given here in Fourier transform)
\begin{equation}
\label{casi6}
\chi^0(\br,\brp,\omega)=\sum_{i}^{\rm occ}\sum_{a}^{\rm virt}\frac{2\omega_{ia}}{\omega_{ia}^2-\omega^2} \phi_i(\br) \phi_a(\br)\phi_i(\brp) \phi_a(\brp),
\end{equation}
where $\phi_i$ and $\phi_a$ are occupied and virtual KS orbitals. In a simpler notation, omitting the integral signs and the variable dependence, practical calculations of the density response are carried out by self consistently solving the following \cite{casi1995}
\begin{equation}
\label{casi6}
\left[ \left( \chi^0 \right)^{-1} - f\right]\delta\rho =\delta \pot{appl},
\end{equation}
with 
\begin{equation}
\label{casi7}
f(\br,\brp,\omega)=\frac{1}{|\br-\brp|}+f_{\rm xc}(\br,\brp,\omega).
\end{equation}
As \eqn{casi6} must hold for any $\delta  \pot{appl}$, comparison with \eqn{casi3a} yields
\begin{equation}
\label{casi8}
\left(\chi \right)^{-1}=\left( \chi^0 \right)^{-1} - f,
\end{equation}
also known as the Dyson equation for the response function \cite{casi1995,gros1996,pete1996}.
\section{Subsystem TD-DFT}
\label{sect_sTD-DFT}
This section is devoted to the derivation of linear-response subsystem TD-DFT. Even though this theory has been first derived by Neugebauer \cite{neug2007}, here it is presented in a different mathematical formalism which makes use of subsystem response functions. The derivations and analyses presented in this section are important as they pave the road to the formalism presented the subsequent sections.
\subsection{Mathematical derivation}
Following the usual decomposition of the density change in subsystem TD-DFT \cite{neug2010a,neug2010,neug2007}, the total electron density change of the system, $\delta \rho$,  due to an external perturbation is given exactly by
\begin{equation}
\label{int3}
\delta\rho(\br,t) = \sumi \delta\rhoi(\br,t),
\end{equation}
where $\delta \rhoi$ is the density change of the single subsystem $I$. 
The subsystem density change is, in all respects, equivalent to a regular TD-DFT density change. For example, one could think that the single subsystem density changes, $\delta\rho_I(\mathbf{r},t)$, may involve inter-subsystem charge transfer type changes so that they may not integrate to zero. In fact, $\int\delta\rho_I(\mathbf{r},t)\d\mathbf{r}=0$ always because one of the defining constraints of the subsystems is that they must be made of a fixed number of electrons. Charge transfer type excitations are naturally accounted for in this theory, as no real-space constraints are imposed on the subsystem densities.

Similarly to \eqn{casi4}, let us consider the effective time-dependent perturbation on subsystem $I$, $\delta \pot{eff}^{I}(\br,t)$, as being a functional of all the subsystem densities, and defined as follows
\begin{equation}
\label{int4}
\delta \pot{eff}^{I}(\br,t)=\underbrace{\delta \pot{appl}(\br,t)}_{\mathrm{perturbation}}+\underbrace{\delta \pot{ind}^{I}(\br,t)}_{\fontsize{8}{10}\selectfont{\begin{array}{c}\mathrm{induced~potential}\\ \mathrm{on~subsystem~I}\end{array}}}.
\end{equation}
Similarly to \eqn{sks}, in the above it is assumed that the applied potential acts on the entire system and therefore it is the same applied potential, $\delta \pot{appl}$, that interacts with all the subsystems.

The induced potential, $\delta \pot{ind}^{I}$, can be defined in terms of functional derivatives of the subsystem KS potential given in \eqn{sks} \cite{neug2010a,neug2007,casi2004}. Defining  
\begin{equation}
K_{IJ}(\br,\brp,t-\tp)=\frac{\delta \pot{ind}^{I}(\br,t)}{\delta \rhoj(\brp,\tp)},
\end{equation}
expressing all quantities in Fourier transform, and applying the convolution theorem, we get
\begin{align}
\label{int5a}
\delta \pot{ind}^{I}(\br,\omega)&=\sumj \int K_{IJ}(\br,\brp,\omega) \delta\rhoj(\brp,\omega) \d\brp  \\
\label{int5b}
K_{IJ}(\br,\brp,\omega)&=\frac{1}{| \br - \brp |} + f_{\rm xc}(\br,\brp,\omega)+ f_{\rm T}(\br,\brp,\omega) - f_{\rm T}^I(\br,\brp,\omega) \delta_{IJ},
\end{align}
where the kinetic kernels, expressed in the time domain, are defined as
\begin{align}
\label{kinkt}
f_{\rm T}(\br,\brp,t-\tp)&=\frac{\delta^2 T_{\rm s}[\rho]}{\delta \rho(\br,t) \delta\rho(\brp,\tp)},\\
\label{kinks}
f_{\rm T}^I(\br,\brp,t-\tp)&=\frac{\delta^2 T_{\rm s}[\rhoi]}{\delta \rhoi(\br,t) \delta\rhoi(\brp,\tp)}.
\end{align}

\eqn{int5b} was derived in Ref.\cite{neug2007} and is found by noticing that $ \frac{\delta^2 T_{\rm s}[\rho]}{\delta \rhor \delta\rhojrp}=\frac{\delta^2 T_{\rm s}[\rho]}{\delta \rhor \delta\rhorp}$ after applying the chain rule. Taking the partial functional derivative with respect to a single subsystem density of functionals of the total supersystem density is equivalent to taking the derivative with respect to the total density \cite{casi2004,neug2007}.

Similarly to \eqn{casi3a} and \eqn{casi3b}, with the aid of the Runge-Gross theorem, the time-dependent subsystem density can be obtained self consistently as 
\begin{align}
\label{int6}
\delta\rhoi(\br,\omega)&=\int \chi_{I}^c (\br,\brp,\omega) \delta \pot{appl} (\brp,\omega)\d\brp  \\
\label{int6a}
                        &= \int \chi_{I}^{0} (\br,\brp,\omega) \delta \pot{eff}^{I} (\brp,\omega)\d\brp,
\end{align}
where $ \chi_{I}^{0}$ is the KS response of the subsystem to the external perturbation, $\chi_I^c$ is the correlated ``coupled'' subsystem response function, and $ \delta \pot{eff}^{I}$ is given by \eqs{int4}{int5a}. 

The above equations hold a great deal of information, e.g.\ the subsystem time-dependent density can be obtained from the simplified subsystem KS response function and the effective time-dependent potential. \eqs{int4}{int5a} can be used in \eqn{int6a}, yielding
\begin{align}
\nonumber
\delta\rhoi(\br,\omega)=&\int  \chi_{I}^{0} (\br,\brp,\omega)  \delta \pot{appl} (\brp,\omega)\d\brp +\\ 
\label{int6b}
&\int \chi_{I}^{0} (\br,\brp,\omega)  \sum_J  K_{IJ}(\br,\brp,\omega) \delta\rhoj(\brpp,\omega) \d\brp\d\brpp.
\end{align}
From the above equation it becomes clear that a subsystem density response is coupled to the density responses of other subsystems through the induced potential \cite{neug2007}. This is a key piece of information, as the subsystem density response will appear in the expressions of all the other subsystem density responses. This is a picture of dynamic coupling between subsystems that reveals how the labeling of the subsystem time-dependent quantities is just a formality. This analysis uncovers the fact that the dynamic response of the supersystem is collective and generally not subsystem additive.

Grouping the terms in \eqn{int6b} involving $\delta\rhoi$ on the lhs and expressing $\delta\rhoi(\br,\omega)$ in terms of integrals of suitable Dirac deltas yields
\begin{align}
\nonumber
&\int \left[ \delta(\br-\brp)\delta(\brp-\brpp)- \chi_{I}^{0} (\br,\brp,\omega) K_{II}(\brp,\brpp,\omega) \right] \delta\rhoi(\brpp,\omega) \d\brp\d\brpp = \\
&=\int  \chi_{I}^{0} (\br,\brp,\omega)  \delta \pot{appl} (\brp,\omega)\d\brp + \int \chi_{I}^{0} (\br,\brp,\omega)  \sum_{J\neq I}K_{IJ}(\brp,\brpp,\omega) \delta\rhoj(\brpp,\omega)   \d\brp\d\brpp.
\label{int6c}
\end{align}
 \eqn{int6c} can be rearranged by acting on the left by $ \left( \chi_{I}^{0}\right)^{-1} (\brppp,\br,\omega)$ and integrating over $\d\br$, using the relation $\int \left(\chi_{I}^{0}\right)^{-1} (\brppp,\br,\omega) \chi_{I}^{0} (\br,\brp,\omega)\d\br=\delta (\brppp-\brp)$,
\begin{align}
\nonumber
&\int \left[ \left(\chi_{I}^{0}\right)^{-1} (\brppp,\brp,\omega)\delta(\brp-\brpp)- \delta(\brppp-\brp)K_{II}(\brp,\brpp,\omega) \right]  \delta\rhoi(\brpp,\omega) \d\brp\d\brpp = \\
&= \delta \pot{appl} (\brppp,\omega) + \int \delta(\brppp-\brp)  \sum_{J\neq I}K_{IJ}(\brp,\brpp,\omega) \delta\rhoj(\brpp,\omega)   \d\brp\d\brpp.
\label{int6d}
\end{align}
After integration over $\brp$ and substitution of $\brppp\to\br$ and $\brpp\to\brp$ the following is obtained
\begin{align}
\label{int6e}
\int \left[ \left(\chi_{I}^{0}\right)^{-1} (\br,\brp,\omega)- K_{II}(\br,\brp,\omega) \right] \delta\rhoi(\brp,\omega) \d\brp &= \delta \pot{appl} (\br,\omega)+\\
\nonumber 
&+ \int  \sum_{J\neq I}K_{IJ}(\br,\brp,\omega) \delta\rhoj(\brp,\omega) \d\brp.
\end{align}
We now define the inverse of the ``uncoupled'' subsystem response function as
\begin{equation}
\label{unc}
\left(\chi_I^u\right)^{-1} (\br,\brp,\omega)= \left(\chi_{I}^{0}\right)^{-1} (\br,\brp,\omega)-K_{II}(\br,\brp,\omega).
\end{equation}
Realizing that \eqn{int6e} holds for \textbf{every} subsystem, the following $N_S \times N_S$ matrix vector equation can be formally constructed
\begin{equation}
\label{coumat}
\bM \boldsymbol\delta \boldsymbol\rho = \boldsymbol{1}\delta \pot{appl},
\end{equation}
with
\begin{equation}
\label{mat}
\bM=\left(
\begin{array}{ccc}
\left(\chi_I^u\right)^{-1} & & -\bK \\
 & \ddots &  \\
-\bK & & \left(\chi_{N_S}^u\right)^{-1} 
\end{array}
\right) ,
\end{equation}
and 
\begin{equation}
\label{vec}
\boldsymbol\delta \boldsymbol\rho = \left(
\begin{array}{c}
\delta \rhoi\\
\vdots \\
\delta \rho_{N_S}
\end{array}
\right),
\end{equation}
where $\bK$ is the matrix composed of the $K_{IJ}$ kernels.
If the matrix in \eqn{mat} is invertible, then the poles of $\bM^{-1}$ occur at the true excitation energies of each subsystem, and hence of the total supersystem.  \eqn{coumat} can be considered the subsystem DFT equivalent of \eqn{casi6}. 
The matrix formulation above yields the coupled subsystem response function defined in \eqn{int6a} as
\begin{equation}
\label{matcou}
\delta\rhoi = \left(\bM^{-1}\right)_{II} \delta \pot{appl},
\end{equation}
thus a formal relationship is
\begin{equation}
\label{matcou2}
\chi_I^c = \left(\bM^{-1}\right)_{II},
\end{equation}
and 
\begin{equation}
\label{matcou3}
\delta\rho = \mathrm{Tr}\left[ \bM^{-1} \right] \delta \pot{appl}.
\end{equation}
Equations (\ref{matcou}--\ref{matcou3}) are well suited to be used in practical calculations. This is because, in practice, all the operators (response functions and kernels) are expressed in a matrix form.  However, in practical calculations, frequency independent kernels (adiabatic approximation) are usually adopted.

It would be very useful to express the above equations completely in terms of subsystem response functions eliminating the $\delta\rho$ and $\pot{appl}$ dependence, as that would lead to a Dyson-type equations formalism relating the time-dependent correlated and KS response of the total system with the ones of the subsystems. The following section provides precisely such a derivation.
\section{Response-Function Formulation of Susbsystem TD-DFT}
\label{sect_sTD-DFT2}
%
Using the definition in \eqn{unc}, \eqn{int6e} can be rearranged as follows
\begin{equation}
\int \left(\chi_I^u\right)^{-1}(\br,\brp,\omega) \delta\rhoi(\brp,\omega) \d\brp = \delta \pot{appl} (\br,\omega) + \int  \sum_{J\neq I} K_{IJ}(\br,\brp,\omega) \delta\rhoj(\brp,\omega) \d\brp,
\end{equation}
and expressing the subsystem density changes in terms of the subsystem response functions and the applied potential [using \eqn{int6}], the above can be simplified to
\begin{align}
\nonumber
&\int \left(\chi_{I}^{u}\right)^{-1} (\br,\brp,\omega) \chi_I^c(\brp,\brpp,\omega)  \delta \pot{appl} (\brpp,\omega) \d\brpp\d\brp =\\
\label{new6e}
&= \int\left[ \delta(\br-\brp) \delta(\brp-\brpp) + \sum_{J\neq I}K_{IJ}(\br,\brp,\omega)\chi_J^c(\brp,\brpp,\omega) \right]  \delta \pot{appl} (\brpp,\omega) \d\brpp \d\brp.
\end{align}
The above equation must hold for any $ \delta \pot{appl} (\brpp,\omega)$, and specifically for $ \delta \pot{appl} (\brpp,\omega) =  \delta(\brpp-\tbr)f(\omega)$, where $f(\omega)$ is any non-zero function of the frequency. Integration over $\d\brpp$ and simplification of the $f(\omega)$ term yields
\begin{equation}
\int \left(\chi_{I}^{u}\right)^{-1} (\br,\brp,\omega) \chi_I^c(\brp,\tbr,\omega) \d\brp =\delta(\br-\tbr) + \int \sum_{J\neq I}K_{IJ}(\br,\brp,\omega)\chi_J^c(\brp,\tbr,\omega) \d\brp.
\end{equation}
Thus, after applying $\chi_{I}^{u} (\brpp,\br,\omega)$ and integration over $\d\br$, the following Dyson-type equation is obtained, in simplified notation,
\begin{equation}
\label{couexact}
\chi_I^c= \chi_I^u + \sum_{J\neq I}^{N_S} \chi_I^u K_{IJ} \chi_J^c.
\end{equation}
The above equation provides a general Dyson equation relating the uncoupled and the coupled subsystem response functions, and where is is clear that the coupling between subsystem responses is mediated by the off-diagonal elements of the kernel matrix $\bK$ which contains exchange-correlation terms as well as kinetic energy terms.

Dyson equations for the response functions involving only the kernels and the KS response functions are derived starting from \eqn{int6e} and read as follows
\begin{align}
\label{uncdyson}
\chi_I^u &= \chi_I^0 +  \chi_I^0 K_{II} \chi_I^u, \\ 
\label{coudyson}
\chi_I^c &= \chi_I^0 +  \chi_I^0 \sumj K_{IJ} \chi_J^c.
\end{align}
Similarly to regular TD-DFT, this formulation shows that the uncoupled response in \eqn{uncdyson} is similar to the one of the isolated subsystem, albeit a small correction in the kernel due to the second functional derivative of the non-additive kinetic energy functional.  

From the above equations, it is evident that if the poles of the response function of subsystem $I$ are well separated from the ones of the other subsystem response functions, then the poles of each subsystem response contain the ones of all other subsystems. This is a particularly interesting result, as it shows that formally the correlated response function of a single subsystem contains information about the electronic spectrum of the entire supersystem. Obviously, in the limit of infinitely separated subsystems, $K_{IJ}(\br,\brp,\omega)$ will be identically zero when $\br$ and $\brp$ span regions of space occupied by different subsystems. Thus, the above observation needs to be taken with a grain of salt as it is valid only if the subsystems are spatially close to each other.

The limiting case of infinite subsystem separation seems to simplify the formalism introducing some degree of subsystem additivity. However, the approximations (such as the adiabatic approximation) usually employed in practical implementations of this theory will likely break down in this limiting case. Retardation effects (finite speed of interactions between subsystems) are completely neglected in practice and it is expected that they will strongly influence the subsystem dynamical coupling when the subsystems are separated by large distances.

Another interesting outcome of this formalism is that when two subsystems have poles at the same frequencies in the isolated case (or in the uncoupled case), then this degeneracy must disappear in the coupled case otherwise the response function would feature an unphysical ``double pole''. This implies that the above formalism is coherent with the existence of Davydov splittings in dimeric systems \cite{neug2007,davydov_excitons}. 
Even though \eqn{couexact} is aesthetically pleasing, it is not suitable for practical calculations. The scheme developed in \eqs{mat}{matcou3} is recovered by rewriting \eqn{couexact} as
\begin{equation}
\label{couexact2}
\chi_I^u= \chi_I^u \left[ \left(\chi_I^u\right)^{-1}\chi_I^c - \sum_{J\neq I}^{N_S} K_{IJ} \chi_J^c \right],
\end{equation}
which leads to
\begin{equation}
\label{couexactmat}
1= \bM \boldsymbol\chi^c,
\end{equation}
where $\bM$ is the same matrix defined in \eqn{mat}. Similarly as before, from \eqn{couexactmat}, the poles of $\bM^{-1}$ are also the poles of the coupled response function. The subsystem TD-DFT equations derived by Neugebauer \cite{neug2007} are readily recovered in this formalism by rewriting the above equation in terms of occupied-virtual KS orbital products and applying the adiabatic approximation (i.e.\ $f_{\rm xc}(\br,\brp,t-\tp)=f_{\rm xc}(\br,\brp,t-\tp)\delta(t-\tp)$, and similarly for the kinetic energy kernels). 
\section{Comparison to TD-DFT of the supersystem}
\label{sect_supra}
Comparison of subsystem DFT with regular KS-DFT is straightforward. In subsystem DFT one has to solve coupled KS-like equations, where the coupling term is conveniently expressed as a potential term, $\pot{emb}$, added to the KS effective potential of the isolated subsystem. This means that the two formalisms involve similar algorithms for practical calculations.  

As it will be clear from the following derivations, this is not the case for the time-dependent extensions. In the following, Dyson-type equations will make it possible to directly compare subsystem DFT with the TD-DFT of the supersystem. The correlated response of the supersystem has a simple relationship to the subsystem correlated responses, taking the functional derivative with respect to $\delta \pot{appl}$ of both sides of \eqn{int3} one obtains
\begin{equation}
\label{stddft}
\chi=\sumi \chi_I.
\end{equation}
Conversely, due to the non-uniqueness of the density partitioning in \eqn{fde1}, the ``simplified'' KS response of the supersystem has no ``simple'' relationship with the subsystem KS responses. 

\subsection{Subsystem versus full KS response function}
In order to find a relationship between the subsystem and the supersystem KS response functions, let us manipulate \eqn{int6} by inverting the subsystem response functions, one by one, in the following manner
\begin{equation}
\label{a1.1}
\left(\chi_I^0\right)^{-1}\delta\rhoi= \delta \pot{eff}^I,
\end{equation}
where we have omitted the integration symbols for sake of a lighter notation. 
An important difference between the induced potential used in subsystem TD-DFT defined in \eqn{int5b} and the one used in the TD-DFT of the supersystem defined in  \eqn{casi5} resides in the kinetic energy kernels. The two can be related,  
\begin{equation}
\label{int14a}
\delta \pot{eff}^I(\br,\omega) = \delta \pot{eff}(\br,\omega) + \delta \pot{T}^I(\br,\omega),
\end{equation}
where, using \eqn{int5a} and \eqn{int5b}, we define the kinetic energy part of the subsystem kernel as 
\begin{align}
\nonumber
\delta \pot{T}^I(\br,\omega) &= \sumj \int \left( f_{\rm T} (\br,\brp\omega)   -  f_{\rm T}^I (\br,\brp\omega) \right) \delta\rho_J(\brp,\omega) \d\brp\\
\label{int14b}
                                    &= \int f_{\rm T} (\br,\brp,\omega) \delta\rho(\brp,\omega) - \int f_{\rm T}^I (\br,\brp,\omega) \delta\rho_I(\brp,\omega),
\end{align}
where \eqn{int3} has been used for the first term of the rhs.
Using \eqn{int14b} in \eqn{a1.1}, we obtain
\begin{equation}
\label{a1.2}
\left[\left(\chi_I^0\right)^{-1}+f_{\rm T}^I \right]\delta\rhoi= \left( 1 + f_{\rm T} \chi^0 \right) \delta \pot{eff},
\end{equation}
where the number $1$ above is intended to be the identity in functional space, i.e.\ a Dirac delta in the position representation. Inverting the operator on the lhs of the above equation and summing over all the subsystems, we obtain the following
\begin{equation}
\label{a1.3}
\sumi \delta\rhoi=\delta\rho = \sumi \left\{ \left[\left(\chi_I^0\right)^{-1}+f_{\rm T}^I \right]^{-1}\left( 1 + f_{\rm T} \chi^0 \right) \right\} \delta \pot{eff}.
\end{equation}
At this point there are several algebraically non-equivalent ways to proceed. Two routes are considered here: the first one leading to an exact expression, and the second one leading to expressions suited for approximations.
\subsubsection{Exact expression}
Extracting $\chi^0$ from the braces of \eqn{a1.3}, and realizing that $\chi^0\delta \pot{eff}=\delta \rho$
\begin{equation}
\label{exa1}
\delta\rho = \underbrace{\sumi \left\{ \left[\left(\chi_I^0\right)^{-1}+f_{\rm T}^I \right]^{-1}\left[ \left(\chi^0\right)^{-1} + f_{\rm T} \right] \right\} \chi^0}_{\mathrm{Identity~operator}} \delta \rho.
\end{equation}
By defining $\left(\chi_I^T\right)^{-1}= \left[\left(\chi_I^0\right)^{-1}+f_{\rm T}^I \right]$, \eqn{exa1} leads to 
\begin{equation}
\label{exa2}
\chi^0= \left[\sumi \left(\chi_I^T\right)^{-1}\right]^{-1} -f_{\rm T}.
\end{equation}

It should be noted that the KS supersystem
considered here is the true KS supersystem. Other formulations of subsystem TD-DFT [31], instead,
considered a supersystem treated with Thomas--Fermi theory.

\subsubsection{Approximate expressions}
A first approximation can be reached directly from \eqn{exa2} in the limit of vanishing kernels, namely
\begin{equation}
\label{exa3}
\left(\chi^0\right)^{-1}= \sumi \left(\chi_I^0\right)^{-1}.
\end{equation}

However, a different approximation can be make by first taking the functional derivative with respect to $ \delta \pot{eff}$ on both sides of \eqn{a1.3}, namely
\begin{equation}
\label{a1.4}
\chi^0 = \sumi \chi_I^0  \left[1 + f_{\rm T}^I \chi_I^0 \right]^{-1} \left( 1 + f_{\rm T} \chi^0 \right),
\end{equation}
which can be arranged to
\begin{equation}
\label{a1.4b}
\chi^0 = \left[ 1 - f_{\rm T} S \right]^{-1} S,
\end{equation}
with $S= \sumi \chi_I^0  \left[1 + f_{\rm T}^I \chi_I^0 \right]^{-1}$.
The above inverse operations expression can be approximated with linear expansions, in the limit of small $f_{\rm T}^I \chi_I^0$ and small $f_{\rm T} \chi^0_I$, to
\begin{equation}
\label{a1.5}
\chi^0 \simeq \sumi \chi^0_I - \sumi \chi_I^0 f_{\rm T}^I \chi_I^0 + \sum_{IJ}^{N_S} \chi^0_I f_{\rm T} \chi^0_J,
\end{equation}
featuring an interesting resemblance to the Dyson equation for the response function. 

%
%
\subsection{Physical meaning of the subsystem KS responses and comparison to PDFT}
\label{sect_pdft}
The derivations in the preceding section stand out as being too complicated for just the KS response, paradoxically in this context known as the ``simplified'' response. What is the significance of such a complicated relationship between the supersystem and the subsystem KS response functions?
What is puzzeling in \eqs{exa2}{a1.5} is that the KS response of the supersystem contains terms coupling KS responses of {\bf different} subsystems. This is not a very good property of this theory, as subsystem additivity is sought in the density, in the correlated response in \eqn{stddft} and it is expected to appear in the KS density response as well. 

This apparent artifact is due to the non-uniqueness and subjectivity of the density partitioning employed in \eqn{fde1}. An indication of this artificial behavior of the subsystem KS responses can be easily shown by considering a more refined version of subsystem DFT known as partition DFT (PDFT). In PDFT  theory \cite{tang2012,elli2010,cohe2009,elli2009} the effective subsystem time-dependent potential is
\begin{equation}
\label{pdft1}
\delta \pot{eff}^I = \delta \pot{appl} + \delta \pot{ind} + \delta \pot{p},
\end{equation}
where $ \delta \pot{p}$ is the change in the partition potential (a quantity shared by all subsystems and thus {\bf unique}). The above equation can be rearranged similarly to the step carried out between \eqn{a1.2} and \eqn{a1.3}, to yield
\begin{equation}
\label{pdft2}
\left[\left( \chi_I^0 \right)^{-1} - f_{\rm p}\right] \delta \rhoi = \left( \chi^0 \right)^{-1}\delta\rho,
\end{equation}
with $f_{\rm p}=\frac{\delta\pot{p}}{\delta\rho}$. The above equation is rearranged to
\begin{equation}
\label{pdft3}
 \chi^0 = \sumi\left[\left( \chi_I^0 \right)^{-1} - f_{\rm p}\right]^{-1}
\end{equation}
which can be approximated assuming small $f_{\rm p}\chi_I^0$ by 
\begin{equation}
\label{pdft4}
\chi^0 \simeq \sumi \chi^0_I - \sumi \chi_I^0 f_{\rm p}^I \chi_I^0.
\end{equation}
The last two equations feature no cross terms coupling the subsystem KS responses. Thus, PDFT provides a more intuitive time-dependent behavior of the subsystems and is completely free of artifacts due to the non-unique partitioning appearing in regular subsystem DFT.

Non-orthogonality also plays a role. For example, a subsystem additive KS response function is expected to be a good approximation to the supersystem KS response function in two limit cases: small electron density overlap between subsystems, and orthogonality between orbitals belonging to different subsystems. This is because in these cases, the non-additive kinetic energy functional is close to being identically zero and the treatment becomes similar to the PDFT case.
\section{Conclusions}
\label{sect_conc}
In this work, the theory of linear-response subsystem TD-DFT is derived in a complete way and analyzed in detail.
For the first time, Dyson equations involving subsystem response functions are derived for linear-response subsystem TD-DFT. Three types of subsystem response functions are considered: coupled, uncoupled and KS. The coupled and uncoupled are exact and approximated correlated subsystem responses, respectively. 

It is found that, for non-infinitely separated subsystems, the pole structure of a correlated (coupled) subsystem response function contains the excitations of the entire supersystem. This  shows that if an applied potential is in resonance with an electronic transition of one subsystem, the electronic response of another subsystem will also be strongly affected. This behavior generally does not fit a picture of ``localized excitations'' but instead is consistent with the idea that the response of a collection of subsystems is collective, and generally delocalized. 

Localization of the excitations may take place whenever the kernel coupling the subsystem's excitations $K_{IJ}$ is small, which is often the case because in practical calculations the subsystems are chosen to be non-bonded molecules. However, a local picture of the time-dependent response of a system is not generally accurate. 

The formalism presented here, specifically the Dyson equations, shows a remarkable similarity with the set of coupled equations one needs to solve for when considering a molecule interacting with a polarizable force field \cite{bene2013}, or the recent model for including non-local correlation in DFT by Tkatchenko and coworkers \cite{tkat2012}. In the two cases mentioned, the local polarizabilities are affected by the presence of other polarizabilities centered on different atoms through Dyson equations, in all respects, similar to \eqn{couexact}. Interestingly, the Hamiltonian that couples polarizabilities is the dipole Hamiltonian, whereas here the full Coulomb kernel is considered.

Another interesting aspect that was uncovered in this work is that, while the correlated response of the supersystem is given by a simple sum of subsystem response functions (subsystem additive), the KS response is not. The kinetic energy kernels are responsible for this non-additivity. Series expansions reveal that the non-additive portions include terms coupling KS responses of different subsystems. This non-additivity is also a feature of another partitioning technique, PDFT. However, in PDFT the non-additive terms do not couple KS responses of different subsystems. This indicates that the unwanted cross-subsystem non-additivity occurring in subsystem TD-DFT is entirely an artifact stemming from the subjective nature of the density partitioning.
\section*{Acknowledgements}
I thank Johannes Neugebauer, Neepa Maitra, Ruslan Kevorkyants, and Henk Eshuis for illuminating discussions. This work is supported by startup funds of the Department of Chemistry and the office of the Dean of FASN of Rutgers University-Newark.

\end{document}